\documentclass[aps,prb,twocolumn,showpacs,superscriptaddress,floatfix]{revtex4-2}  
\usepackage{float}
\usepackage{ucs}
\usepackage{natbib}
\usepackage{graphicx}  
\usepackage{bm}        
\usepackage{amssymb}   
\usepackage{amsmath}
\usepackage{color}
\usepackage{CJK}
\usepackage{times}
\usepackage{verbatim}
\usepackage{mathrsfs}
\usepackage{hyperref}
\usepackage{ulem}
\usepackage{physics}
\hypersetup{colorlinks = True, urlcolor=blue, linkcolor=blue, citecolor=blue}

\usepackage{cancel}
\usepackage{comment}

\begin{document}
\title{Machine learning topological energy braiding of non-Bloch bands}

\author{Shuwei Shi}
\affiliation{School of Physics and Electronic Engineering, Jiangsu University, Zhenjiang, Jiangsu 212013, People’s Republic of China}
\affiliation{Jiangsu Engineering Research Center on Quantum Perception and Intelligent Detection of Agricultural Information, Zhenjiang 212013, People’s Republic of China}

\author{Shibing Chu}\altaffiliation{c@ujs.edu.cn}
\affiliation{School of Physics and Electronic Engineering, Jiangsu University, Zhenjiang, Jiangsu 212013, People’s Republic of China}
\affiliation{Jiangsu Engineering Research Center on Quantum Perception and Intelligent Detection of Agricultural Information, Zhenjiang 212013, People’s Republic of China}

\author{Yuee Xie}
\affiliation{School of Physics and Electronic Engineering, Jiangsu University, Zhenjiang, Jiangsu 212013, People’s Republic of China}
\affiliation{Jiangsu Engineering Research Center on Quantum Perception and Intelligent Detection of Agricultural Information, Zhenjiang 212013, People’s Republic of China}

\author{Yuanping Chen} \altaffiliation{ypchen@ujs.edu.cn}
\affiliation{School of Physics and Electronic Engineering, Jiangsu University, Zhenjiang, Jiangsu 212013, People’s Republic of China}
\affiliation{Jiangsu Engineering Research Center on Quantum Perception and Intelligent Detection of Agricultural Information, Zhenjiang 212013, People’s Republic of China}
\date{\today}
\begin{abstract}
Machine learning has been used to identify phase transitions in a variety of physical systems. However, there is still a lack of relevant research on non-Bloch energy braiding in non-Hermitian systems. In this work, we study non-Bloch energy braiding in one-dimensional non-Hermitian systems using unsupervised and supervised methods. In unsupervised learning, we use diffusion maps to successfully identify non-Bloch energy braiding without any prior knowledge and combine it with k-means to cluster different topological elements into clusters, such as Unlink and Hopf link. In supervised learning, we train a Convolutional Neural Network (CNN) based on Bloch energy data to predict not only Bloch energy braiding but also non-Bloch energy braiding with an accuracy approaching $100\%$. By analysing the CNN, we can ascertain that the network has successfully acquired the ability to recognise the braiding topology of the energy bands. The present study demonstrates the considerable potential of machine learning in the identification of non-Hermitian topological phases and energy braiding.
\end{abstract}

\maketitle

\section{Introduction}
In nature, braids and knots are prevalent structural forms across various fields\cite{adams1994knot}. These concepts are often employed to describe topological phase transitions, where systems with braiding structures exhibit rich and stable topological properties. 
Unlike Hermitian systems, non-Hermitian systems feature complex eigenenergies, leading to unique phenomena such as exceptional points\cite{kawabata2019classification,rui2019topology,zhang2020non,xiao2021observation,sayyad2023symmetry,BingBingWang} and non-Hermitian skin effect\cite{yao2018edge,song2019non,li2020critical,zhang2021acoustic,liang2022dynamic,kawabata2023entanglement,zhou2023observation}. Studies have shown that the complex energy bands in one-dimensional non-Hermitian systems can form intricate and diverse braiding topologies, which are classified through the conjugacy classes of braid groups\cite{shen2018topological,kawabata2019symmetry,wojcik2020homotopy,li2021homotopical,Wang20211,wang2021generating,zhang2021acoustic,patil2022measuring,zhang2023observation}.
These complex braiding structures have been experimentally verified on platforms including photonic systems\cite{science.aaa8685,Wang20211}, acoustic systems\cite{zhang2023observation,qiu2023minimal,zhang2023experimental} and quantum circuits\cite{yu2022experimental}. However, identifying complex energy braiding\cite{chen2024machine}, especially non-Bloch braiding\cite{li2022topological}, remains challenging. An efficient identification method is urgently needed.

Machine learning methods have rapidly developed and found applications in fields such as image recognition\cite{7780459,app132312649}, natural language processing\cite{science.aaa8685}, recommender systems\cite{1423975}, and motion data analysis\cite{3907002}. Their powerful ability to analyze large-scale data and uncover hidden patterns has led to advances in condensed matter physics, quantum physics, and astrophysics\cite{RevModPhys.91.045002}. In physics, supervised learning has been used to identify phase transitions and topological invariants\cite{PhysRevB.103.035413,PhysRevA.103.012419}. 
Unsupervised methods, such as diffusion maps combined with k-means, can identify phase transition points and Bloch braiding quickly and in a purely data-driven manner without prior physical knowledge\cite{COIFMAN20065,PhysRevB.102.134213,PhysRevLett.126.240402,Yu2021ExperimentalUL,Long2024,chen2024machine}. 
However, most current research has focused on Bloch bands under periodic boundary conditions, yielding idealized results. Non-Hermitian systems under open boundary conditions are more representative of realistic systems, exhibiting phenomena such as the non-Hermitian skin effect, where eigenenergies collapse. This disrupts the topological braids of the Bloch bands and invalidates the topological invariants defined in the Brillouin zone, making non-Bloch braiding identification more challenging\cite{li2022topological}.

In this work, we utilize diffusion maps and Convolutional Neural Networks (CNN) to predict energy braiding in non-Hermitian systems. We employ diffusion maps to identify non-Bloch braiding without prior knowledge and train a CNN to predict Bloch energy braiding, achieving high accuracy even for topological elements not included in the training set. Our study demonstrates the potential of machine learning for studying phase transitions and energy braiding in non-Hermitian systems.
\section{algorithms}
\subsection{unsupervised algorithm}
When trying to identify topological phase transitions in non-Hermitian systems, linear downscaling methods (e.g., PCA) may not be effective in capturing the topology structure because they assume that the main direction of change in the data is linear. In contrast, nonlinear dimensionality reduction methods (e.g., diffusion maps) can capture the topology structure by constructing a neighbourhood map that preserves this nonlinear structure in the low-dimensional space\cite{COIFMAN20065}.

For example, consider a two-band non-Hermitian system with the Hamiltonian expressed as:
\begin{eqnarray}
	\begin{aligned}
		H(k)&=\boldsymbol{d}({k})\hat{\boldsymbol{\sigma}}=d_x(k)\hat{\sigma}_x+d_y(k)\hat{\sigma}_y+d_z(k)\hat{\sigma}_z,
	\end{aligned}
\end{eqnarray}
where $\hat{\sigma}_{x,y,z}$ are the Pauli  matrices and $k$ is the Bloch wave vector.
We can construct such a dataset ${\boldsymbol X } = \{{ \boldsymbol{x}_{1}},\,{ \boldsymbol x_{2}},\,\cdots { \boldsymbol x_{l}} \}$, where ${ \boldsymbol x_i=\mathbf d^i{(k)}/(E^i_+(k)-E^i_-(k))}$ representing the $i$-th normalized Hamiltonian sample with $E_\pm(k)=\pm \sqrt{d_x^2(k)+d_y^2(k)+d_z^2(k)}$ .

The distance between $\boldsymbol x_i$ and $\boldsymbol x_j$ is defined as: $\boldsymbol M_{i,j}=\|\boldsymbol x_i-\boldsymbol x_j\|_p^{2}$. In recent research, there has been discussion regarding the utilization of $p=1,\, 2, \infty$ cases in unsupervised clustering topological phases\cite{PhysRevB.102.134213} . In this paper, we set $p=1$, unless otherwise specified. The similarity matrix is derived from the Gaussian kernel 
\begin{eqnarray}
	\boldsymbol A_{i,j}=\exp\left(-\frac{\boldsymbol M_{i,j}}{2\epsilon N^2}\right),
\end{eqnarray}
where $\epsilon$ is the Gaussian kernel coefficient($0<\epsilon \ll 1$). $N$ is an adjustable parameter related to the specific model.
$\boldsymbol A_{i,j}\in\left[0,1\right]$ represents the similarity between $\boldsymbol x_i$ and $\boldsymbol x_j$. The probability transition matrix
is defined as $\boldsymbol P_{i,j}=\boldsymbol A_{i,j}/\sum_{j=1}^{N}\boldsymbol A_{i,j}$. After $2t$ steps of random walk, we can use diffusion distance to describe the connectivity between ${\boldsymbol x_i}$ and $\boldsymbol x_j$
\begin{equation}
	\begin{aligned}
		\boldsymbol D_{t}(i,j)=\sum_{k=1}^l\frac{\left(\boldsymbol{P}_{i,k}^t-\boldsymbol{P}_{j,k}^t\right)^2}{\sum_q\boldsymbol{A}_{k,q}}
		&= \sum_{k=1}^{l-1}\lambda_{k}^{2t}[(\psi_k)_i - (\psi_k)_{j}]^2,
	\end{aligned}
\end{equation}
where $\lambda_k$ and $\psi_k$ are the $k$-th eigenvalue and right eigenvector of $\boldsymbol P$, respectively.

After prolonged diffusion, only the few components $\psi_k$ with $\lambda_k\approx1$ dominate the diffusion process. Therefore, the manifold diffusion distance information of the samples is encoded in these components, thus enabling the downscaling of high dimensional data\cite{COIFMAN20065}. By further applying clustering algorithms (e.g., k-means) to low-dimensional data, it is possible to cluster different topological phases without prior knowledge. Previous research has demonstrated that the number of $\lambda_k\approx1$ corresponds to the the number of topological clusters\cite{PhysRevB.102.134213}. 

\subsection{supervised algorithm}
\begin{figure}[H]
	\centering
	\includegraphics[width=0.45\textwidth]{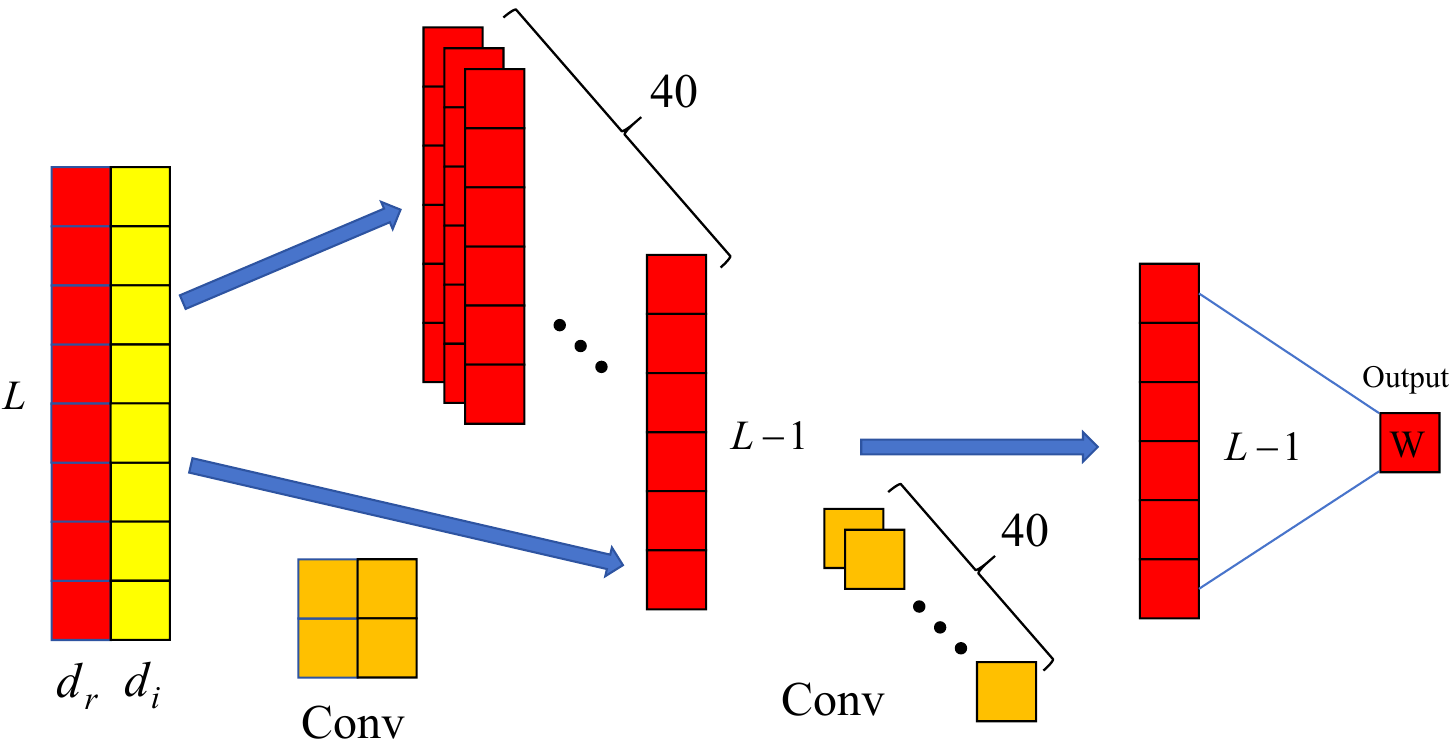}
	\caption{Schematic of the CNN. Here $d_r$ and $d_i$ represent the real and imaginary parts of the input data, respectively.}
	\label{figcnn}
\end{figure}

For supervised learning, we use a CNN as shown in Fig.~\ref{figcnn}.

\section{model and results}
\subsection{unsupervised learning results}
\begin{figure*}
	\centering
	\includegraphics[width=0.9\textwidth]{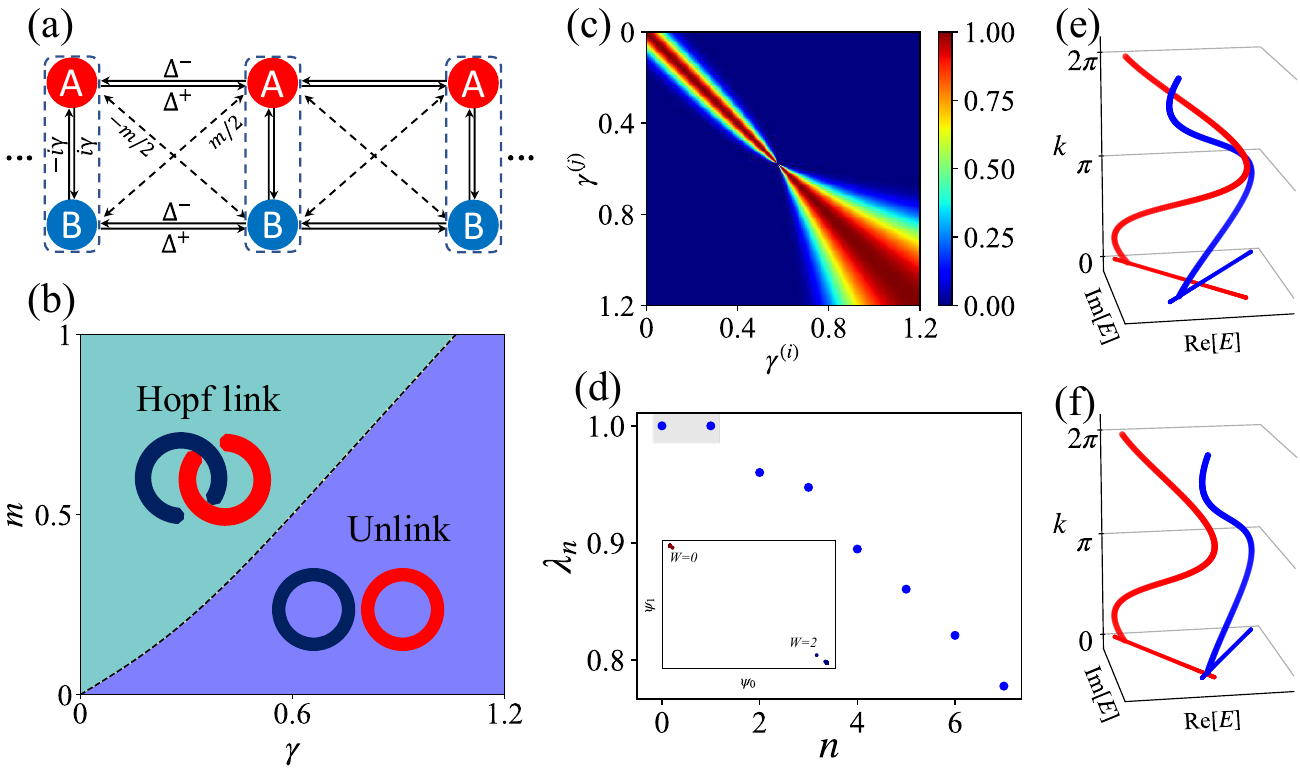}
	\caption{(a) Sketch of the two bands non-Hermitian lattice model. (b) Phase diagram of the non-Hermitian system under open boundary condition with $\Delta=1.0$ and $\delta=0.7$. (c) The visualisation of the Gaussian kernel. The breakpoint indicates the predicted phase transition point $\gamma\approx0.58$. (d) The first eight eigenvalues of the diffusion matrix $\boldsymbol P$. The inset shows the result of k-means. The first two largest eigenvalues correspond to the two topological clusters in the inset. The samples are generated uniformly of $\gamma \in [0,1.2]$ with $\Delta=1$, $\delta=0.7$ and $m=0.5$.  The Gaussian kernel coefficient $\epsilon=0.01$. (e-f) Complex energy braiding of non-Bloch bands in $(\mathrm{Re}[E],\mathrm{Im}[E],k)$ space. (e) and (f) are topologically distinct and correspond to Hopf link and Unlink, respectively.}
	\label{NHOBC}
\end{figure*}

Let us begin with a two-band non-Hermitian lattice as shown in Fig.~\ref{NHOBC}(a). The Hamiltonian of this lattice model takes the form
\begin{eqnarray}
	&& H =\sum_{n} \sum_{\sigma\in \{A,B\}} \left(\Delta^- c_{n,\sigma}^{\dag} c_{n+1,\sigma}+\Delta^+ c_{n+1,\sigma}^{\dag} c_{n,\sigma}\right) \nonumber\\
	&& \quad \quad  + \frac{m}{2}\left( c_{n+1,A}^{\dag}c_{n,B}-c_{n+1,B}^{\dag}c_{n,A}+H.c.\right) \nonumber\\
	&& \quad \quad - \left( i\gamma c_{n,A}^{\dag}c_{n,B}+H.c. \right),
	\label{Hamilt}
\end{eqnarray}
here $c^{\dag}_{n, \sigma} \, (c_{n, \sigma})$ denotes the creation (annihilation) operator of the sublattice $\sigma \in \{A,B\}$ on the $n$th site. The terms $\Delta^{\pm} = (\Delta \pm \delta)/2$ represent the non-reciprocal nearest-neighbour hopping within the same sublattice. The nonzero parameter $\delta$ breaks the Hermicity of the model. The parameter $m$ characterizes the reciprocal nearest-neighbor hopping between distinct sublattices, while $\gamma$ denotes the intracell coupling. All parameters are real\cite{li2022topological}.



The form of the real-space Hamiltonian expressed by Eq.(\ref{Hamilt}) can be written in the form of the Bloch Hamiltonian in momentum space as
$H(k) = \Delta \cos k -i \delta \sin k + (\gamma + m \sin k) \hat\sigma_{y}$.
We can obtain the non-Bloch Hamiltonian $H(\beta)$ by extending the BZ into GBZ with $\beta=re^{ik}, r = \sqrt{\left|{(\Delta + \delta + im)}/{(\Delta - \delta - im)}\right|}, k\in\mathbb{R}$. The energies of the non-Bloch
bands take the form $E_\pm(\beta)=\pm \gamma+0.5(\Delta-\delta\mp im)\beta+0.5(\Delta+\delta\pm im)\beta^{-1}$.
The non-Bloch winding number is defined as 
\begin{eqnarray}
	W=\frac1{2\pi i}\oint_{C_\beta}d\ln\det\left[H(\beta)-\frac12\mathrm{Tr}[H(\beta)]\right]\label{W},
\end{eqnarray}
which describes the braid degree of two non-Bloch bands\cite{li2022topological}.

In order to explore the non-Bloch braiding in this model as shown in Fig.~\ref{NHOBC}(b), we consider a dataset ${\boldsymbol X } = \{{ \boldsymbol{x}_{1}},\,{ \boldsymbol x_{2}},\,\cdots { \boldsymbol x_{l}} \}$ with $m=0.5$, $\Delta=1$, $\delta=0.7$ and variable
$\gamma \in \left[0,1.2\right]$, where ${ \boldsymbol x_i=\mathbf d^i{(\beta)}/(E^i_+(\beta)-E^i_-(\beta))}$. The results of unsupervised learning are shown in Fig.~\ref{NHOBC}(c-d). From Fig.~\ref{NHOBC}(c), the similarity matrix is separated into two blocks, which corresponding to the two largest eigenvalues $\lambda\approx1$ of the diffusion matrix shown in Fig.~\ref{NHOBC}(d). As a result, the diffusion maps identifies two distinct topological phases. In addition, we can get the phase transition point $\gamma\approx0.58$ form Fig.~\ref{NHOBC}(c), which is consistent with the theoretical value $\gamma_c=m(r^2+1)/2r\approx0.579$.

We compare the results of unsupervised learning with those calculated in Eq.~\ref{W}. For $\gamma<\gamma_c$, $W=2$, corresponding to Hopf link, and for $\gamma>\gamma_c$, $W=0$, corresponding to Unlink, as shown in Fig.~\ref{NHOBC}(b). In Fig.~\ref{NHOBC}(e-f), we display the braiding of non-Bloch bands in (Re$[E]$,Im$[E]$,$k$) space with $\gamma=0.3$ and $\gamma=0.8$, corresponding to the blue and red topological clusters in the inset of Fig.~\ref{NHOBC}(d), respectively.  

\subsection{supervised and transfer learning results}
\subsubsection{learning Bloch braiding}
\begin{figure}
	\centering
	\includegraphics[width=0.45\textwidth]{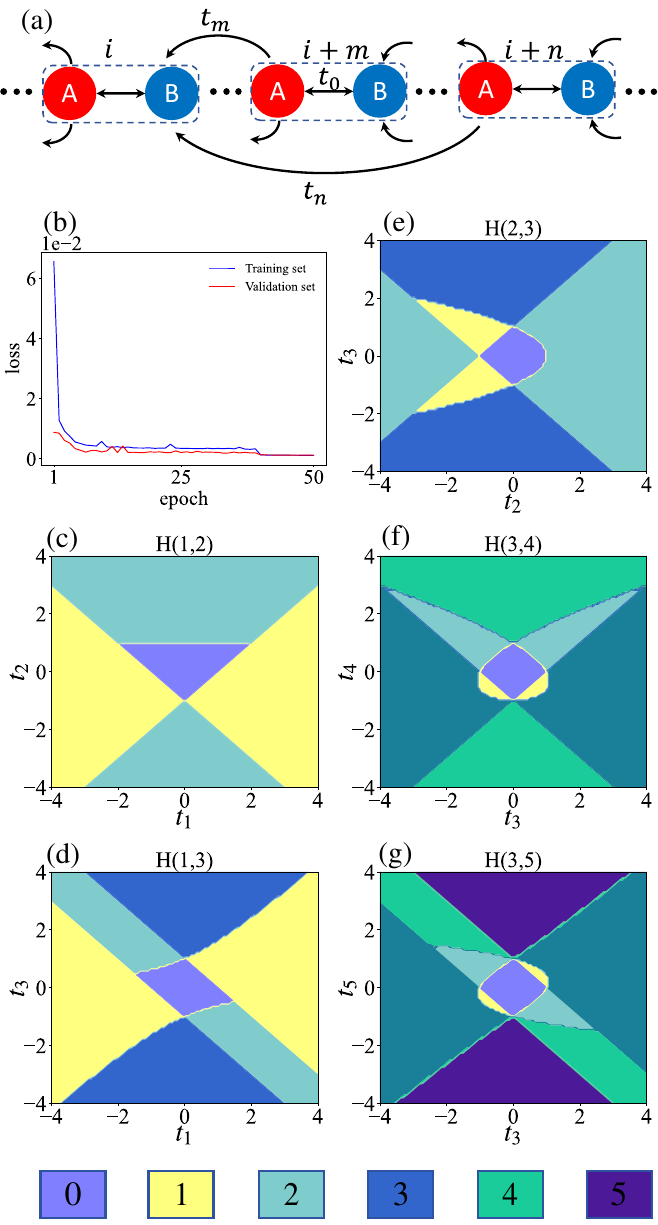}
	\caption{(a) Sketch of the two-band non-Hermitian lattice model $H_{(m,n)}$. (b) The loss with each training and validation. (c-g) The predicted phase diagrams, along with the corresponding braid degrees across different parameter spaces.}
	\label{lll1}
\end{figure}

In this section, we consider a two-band non-Hermitian lattice as shown in Fig.~\ref{lll1}(a). The Hamiltonian of this lattice model takes the form
\begin{equation}
	{H}_{(m,n)}=\begin{pmatrix}0&t_0\\t_0+t_me^{imk}+t_ne^{ink}&0\end{pmatrix},
\end{equation}
where $t_0$ is reciprocal intracell coupling. $t_m$ and $t_n$ are nonreciprocal intercell hoppings that spans $m$ and $n$ lattices, respectively\cite{zhang2023observation}. 

The complex energy spectra reads $E_\pm(k)=\pm\sqrt{t_0(t_0+t_me^{imk}+t_ne^{ik})}$. For any two energy bands satisfy the separable band condition($E_+(k)\neq E_-(k)$ for all $k\in[0,2\pi]$), a topological invariant $W$ can be defined to classify their braiding topology\cite{zhang2023observation}, 
\begin{equation}
	W=\frac{1}{2\pi i}\oint_0^{2\pi}\left(\frac{d\ln E_+}{dk}+\frac{d\ln E_-}{dk}\right)dk.
	\label{w}
\end{equation}

We trained a generic CNN as shown in Fig.~\ref{figcnn} to predict the Bloch energy braiding in this model. The CNN in our research has two convolution layers with 40 kernels of size ($1,2,2$) and 1 kernel of size ($40,1,1$), followed by a fully connected layer before the output layer. Similar to but different from previous studies\cite{zhang2018machine,zhang2021machine}, only the first convolutional layer is followed by the nonlinear activation function ReLU. We find that the prediction accuracy of the CNN depends on the first convolutional layer and its activation function. Shallow networks are more effective than deep networks in learning braid degree. Specifically, simply increasing the number of layers of the CNN can instead cause it to lose its generalisation ability. 

The input data are normalised spectral correlation configurations $\mathbf{d}(n)=\{\mathrm{Re}[E^2(2\pi n/L)],\mathrm{Im}[E^2(2\pi n/L)]\}$, $E=E_+(k)-E_-(k)$. In the following, we set $L = 201$, which is large enough for the CNN to capture the topology of the complex energy bands.

The labels corresponding to each configuration are calculated by the Eq.~\ref{w}. The output of the CNN is a real number $w$ representing the braid degree. The objective function to be optimized is defined by 
\begin{eqnarray}
	J=\frac1N\sum_{i=1}^N({w}_i-W_i)^2,
\end{eqnarray}
where $w_i$ and $W_i$ are, respectively, the predicted and true braid degree of $i$th sample, and $N$ represents the size of the training data set. We uniformly sample $9\times10^{4}$ configurations from ${H}_{(1,2)}$ and ${H}_{(1,3)}$, containing Unlink$(W=0)$, Unknot$(W=1)$, Hopf link$(W=2)$ and Trefoil$(W=3)$, respectively. There are five different types of test data,  (\romannumeral1,\romannumeral2) two test sets with $1\times10^{4}$ configurations generated from ${H}_{(1,2)}$ and ${H}_{(1,3)}$, respectively, (\romannumeral3,\romannumeral4,\romannumeral5) $1\times10^{4}$ configurations sampled from ${H}_{(2,3)}$, ${H}_{(3,4)}$ and ${H}_{(3,5)}$, respectively. The batch size for training is set to 50 and the epoch is set to 50.

After training, the losses on the training and validation sets are shown in Fig.~\ref{lll1}(b). It can be observed that the neural network converges very quickly and the losses stay at a low level and no overfitting occurs. 
We test with the test data and round the output to plot the phase diagram of the predictions on the different test sets as shown in Fig.~\ref{lll1}(c-g).  
In Fig.~\ref{lll1}(c) and (d), the CNN successfully predicts the phase diagram using the test sets (\romannumeral1) and (\romannumeral2), which seems natural since our training data are also sampled with the same parameters. However, we can verify the effectiveness of the CNN in Fig.~\ref{lll1}(e-g). In Fig.~\ref{lll1}(e), the CNN successfully predicts the braid degrees under the ${H}_{(2,3)}$ case, which are contained in the training set. In Fig.~\ref{lll1}(f) and (g), the CNN also predicts the braid phases with high accuracy, even predicting two braid degrees, $W=4$ and $W=5$, that are not included in the training set. At this point, we have demonstrated the ability of the CNN to predict the braid degree under different parameter spaces of the same model.

\subsubsection{transfer learning non-Bloch braiding}
\begin{figure}
	\centering
	\includegraphics[width=0.45\textwidth]{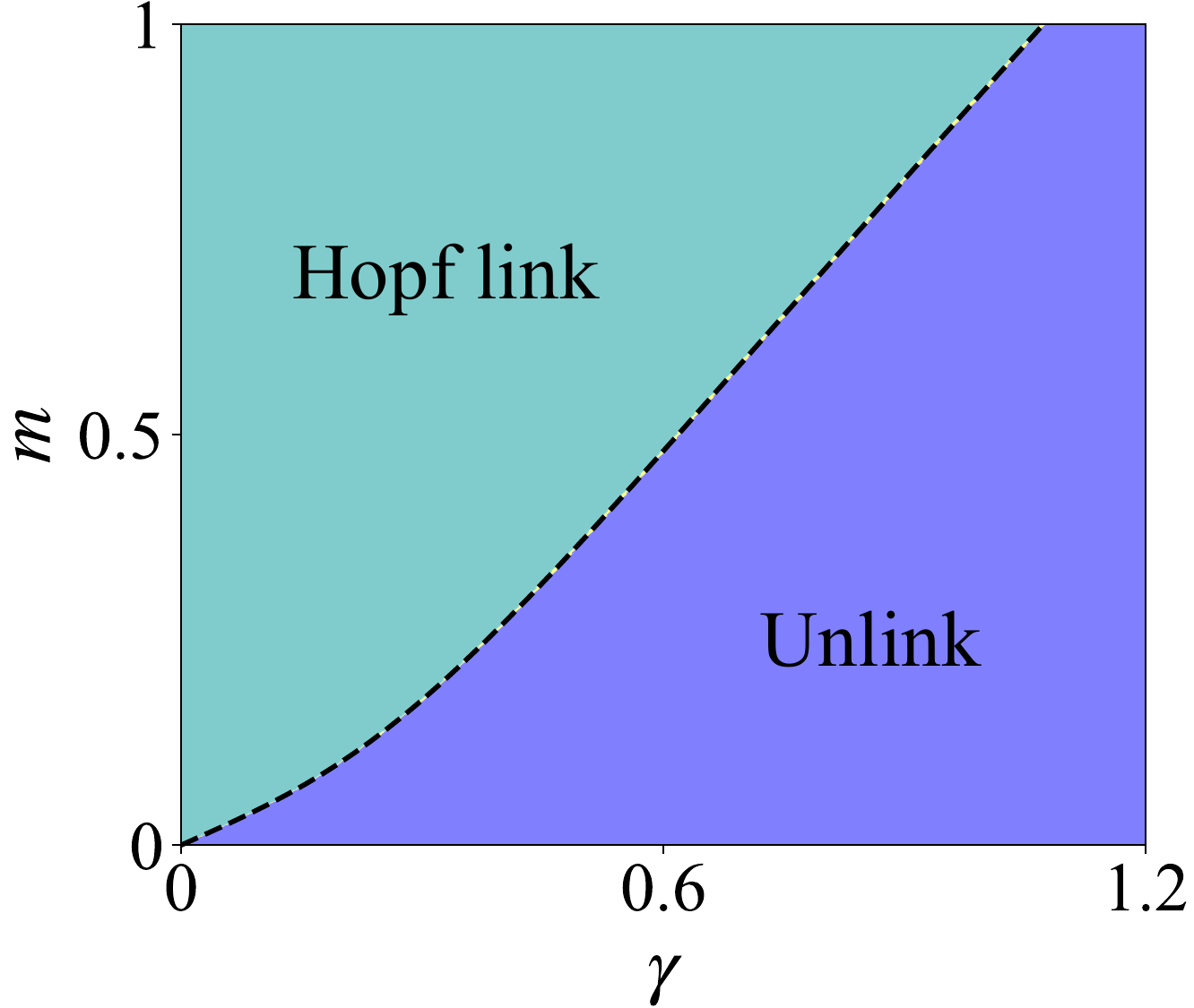}
	\caption{The phase diagram predicted by the pre-trained CNN in the $\gamma-m$ plane with $\Delta=1.0$ and $\delta=0.9$.}
	\label{lll2}
\end{figure}

Further, we use the above pre-trained CNN to recognise non-Bloch braiding in the non-Hermitian system as shown in Fig.~\ref{NHOBC}(a). 
The input data can be chosen as $\mathbf{d}(n)=\{\mathrm{Re}[E^2(2\pi n/L)],\mathrm{Im}[E^2(2\pi n/L)]\}$, $E=E_+(\beta)-E_-(\beta)$ with $\Delta=1.0$ and $\delta=0.9$, varying $\gamma \in [0,1.2]$ and $m \in [0,1]$. 
The predicted phase diagram with respect to the parameters $\gamma$ and $m$ is shown in Fig.~\ref{lll2}. The dashed line is the theoretical value of the phase boundary. Without retraining, the CNN almost perfectly predicts the presence of Unlink and Hopf link elements accompanied by the exact phase boundary in the phase diagram.

To further understand the workflow of the CNN, we analyse the training details.
The first layer of the CNN is a convolutional layer with 40 convolutional kernels to perform convolutional operations on the input data,
\begin{eqnarray}
	\begin{aligned}
		{B}^i(n)=\boldsymbol f [&A_{11}^i {\boldsymbol d}_r(2\pi(n-1)/L)+A_{12}^i {\boldsymbol d}_i(2\pi(n-1)/L)\\
		&+A_{21}^i {\boldsymbol d}_r(2\pi n/L)+A_{22}^i {\boldsymbol d}_i(2\pi n/L)+A_0^i ],
		\label{B}
	\end{aligned}
\end{eqnarray}
where $A$ is a kernel of size $2\times2$, $i=1, ...,40$, $n=1,...,L$ and $\boldsymbol f(x)$ is the activation function.
The second layer is a convolutional layer without  an activation function that takes the form,
\begin{eqnarray}
	\begin{aligned}
		D^i(n)=\sum_{i=1}^{40}C^iB^i(n)+C_0^i,
	\end{aligned}
\end{eqnarray}
where $C^i$ is the $1\times1$ kernel. $D^i(n)$ depend on the winding angle of the two bands.
Finally, all 40 neurons are mapped to a single output neuron to obtain the braid degree,
\begin{eqnarray}
	\begin{aligned}
		W=\sum_{n=1}^{L} M_{n}D(n)+M_0.
	\end{aligned}
\end{eqnarray}

During the training process, the neural network successfully fit all the parameters above. Through the above analysis, we find that the Eq.~\ref{B} used to extract the braid information is universal, and then the output of the Eq.~\ref{B} can be linearly summed to obtain the braid degree.
\section{Conclusion}

In summary, we applied unsupervised and supervised methods to identify the energy braiding in non-Hermitian systems. The unsupervised learning algorithm, diffusion maps, can identify the phase transition points and separate different braiding elements into clusters.  
In supervised learning, we trained a CNN to capture the braiding topology of complex energy in 
non-Hermitian systems. It is worth noting that the CNN trained under Bloch conditions can predict not only Bloch braiding but also non-Bloch braiding. Our methods can be extended to identify other knots and links in non-Bloch bands. Our study demonstrates the potential of machine learning in identifying non-Hermitian topological phases and complex energy braiding.
\section{Data availability statement}
The data that support the findings of this study are available
upon reasonable request from the authors.
\section{ACKNOWLEDGMENTS}
This work was supported by the National Natural Science 
Foundation of China (No. 12074150, 12174157 and 11904137).
\normalem
\bibliography{MLbraiding.bib}
	
\end{document}